# Evidence for speed-dependent effects in NH$_3$ self-broadened spectra: towards a new determination of the Boltzmann constant

M. Triki, C. Lemarchand, B. Darquié, P. L. T. Sow, V. Roncin, C. Chardonnet and C. Daussy[*]

*Laboratoire de Physique des Lasers, UMR 7538 CNRS, Université Paris 13, 99 av. J.-B. Clément, 93430 Villetaneuse, France*

E-mail: christophe.daussy@univ-paris13.fr

**Abstract:**

In this paper we present an accurate analysis of the shape of an isolated rovibrational ammonia line from the strong ν$_2$ band around 10 μm, recorded by laser absorption spectroscopy. Experimental spectra obtained under controlled temperature and pressure, are confronted to various models that take into account Dicke narrowing or speed-dependent effects. Our results show clear evidence for speed-dependent broadening and shifting, which had never been demonstrated so far in NH$_3$. Accurate lineshape parameters of the ν$_2$ saQ(6,3) line are obtained. Our current project aiming at measuring the Boltzmann constant, $k_B$, by laser spectroscopy will straight away benefit from such knowledge. We anticipate that a first optical determination of $k_B$ with a competitive uncertainty of a few ppm is now reachable.

## I. INTRODUCTION

NH$_3$ is a very important molecule for interstellar and planetary observations [1-3]. It is the main molecule observable in a variety of astrophysical environments and it is present in the interstellar medium, in giant planets of solar system (like Jupiter and Saturn), in brown dwarfs, and in cometary coma. Ammonia is also present in trace quantities in the Earth's atmosphere and has been detected in many industrial sites. Absolute measurement of ammonia concentration in air is thus a real challenge for environmental and human health issues [4-6]. With its large dipolar momentum and its inversion spectrum, it is also a very important molecule for fundamental infrared and microwave (MW) spectroscopy, metrological applications or fundamental physics tests. It is for instance a very good candidate for an accurate test of extensions of the Standard Model thanks to its inversion transitions particularly sensitive to variation of the electron-to-proton mass ratio [7-9].

---

[*] Author, to whom any correspondence should be addressed.

In this paper, we focus on the lineshape analysis of a single isolated rovibrational absorption line ammonia. Under controlled temperature and pressure, absorption profiles provide a probe of intermolecular interactions that directly affect the molecular dynamics. Accurate knowledge of the $NH_3$ spectrum – especially the strong $\nu_2$ band in the 10 µm region – as well as lineshape parameters are in particular indispensable to enable astrophysical and planetary interpretations. Several accurate laboratory studies have been performed, using high resolution Fourier transform or tunable laser diode spectrometers, to provide lineshape parameters for the $\nu_2$ band [10-18]. Absolute frequencies, intensities, self-broadening and self-shift coefficients have been recently re-measured in the R branch of the $\nu_2$ band by Guinet *et al.* [19]. The authors bring to light large discrepancies between values recommended by HITRAN08 and recent results published by several groups (about 10% difference for the self-broadening) [20]. All of these analyses have been performed for pressures ranging typically from 10 mbar to 1 bar and by fitting a Voigt profile to experimental absorption lines (including line mixing effects). However in 1993 Pine *et al.* have observed that $NH_3$ lineshapes exhibit significant deviations from the Voigt profile in the $\nu_1$ band. Over the covered pressure range, it was not possible to reliably distinguish between Dicke narrowing and speed-dependent effects [21, 22]. We present here an accurate lineshape analysis of a self-broadened rovibrational absorption transition of ammonia in the strong $\nu_2$ band, recorded by linear absorption laser spectroscopy. Spectra are recorded at very low pressures (below 20 Pa) and analyzed using various models (which include Dicke narrowing or speed-dependent effects) which enables to clearly determine the profile which best matches the data. Accurate lineshape parameters are then deduced from a least-squares fit of the data.

In the last section of this paper we exploit the latter results in the frame of our project dedicated to the Boltzmann constant measurement by laser spectroscopy [23, 24]. In this experiment the main source of uncertainty comes from the modelisation of collisional effects which directly affects the absorption lineshape used for analysis. We present here an accurate determination of the impact of the absorption profile parameters' uncertainties on that of the Boltzmann constant. We conclude that the determination of the Boltzmann constant, $k_B$, by laser spectroscopy with a competitive uncertainty of a few ppm is now reachable.

## II. THE SPECTROMETER

Experimental absorption lineshapes have been recorded using the $CO_2$ laser based spectrometer presented in FIG. 1 and described in details in ref. [25-27]. The $CO_2$ laser source is coupled to a MW electro-optic modulator (EOM) which generates tunable sidebands, from 8 to 18 GHz, on both sides of the fixed laser frequency. After the MW EOM, a grid polarizer and a Fabry-Perot cavity (FPC) are used to filter out the residual carrier and the unwanted sideband. The FPC is also used to stabilize the intensity of the transmitted reference beam (A). The probe beam (B) feeds an ammonia absorption cell for spectroscopy. The absorption length of the cell can be adjusted from 37 cm in a single pass configuration (SPC) to 3.5 m in a multi-pass configuration (MPC). Both reference (A) and probe (B) beams are amplitude-modulated at f=40 kHz via the 8-18 GHz EOM for noise filtering, after demodulation of the signals at f in a lock-in amplifier. The sideband is tuned close to the molecular resonance and scanned to record the Doppler profile.

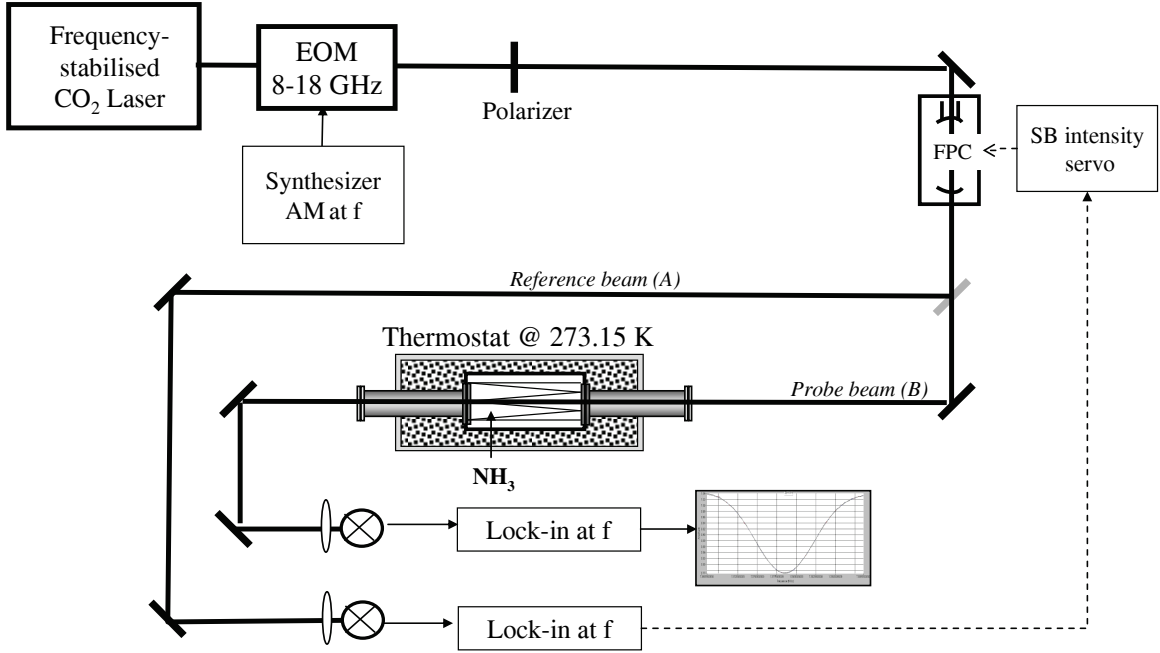

FIG. 1. Experimental setup (AM: Amplitude Modulation, EOM: Electro-Optic Modulator, FPC: Fabry Perot Cavity, Lock-in: Lock-in Amplifier).

The absorption cell is pumped down to a residual pressure of about $10^{-3}$ Pa, 3 to 4 orders of magnitude below the $NH_3$ filling pressure, ranging from 1 to 20 Pa. It is measured by a capacitive pressure gauge (MKS Baratron type 390) with 1% uncertainty. To avoid sample pollution, due to outgasing of the absorption cell during the experiment, spectra are recorded during short series of 30 min after which the absorption cell is pumped before a new acquisition. The absorption cell is placed inside a copper thermal shield which is itself inside a stainless steel enclosure immersed in a thermostat filled with an ice-water mixture which stabilizes the cell's temperature close to 273.15 K. The temperature is measured with small 25 $\Omega$ Hart capsule platinum resistance thermometers placed on the absorption cell. The thermostat and the temperature chain are described in details in reference [27]. Thanks to this set-up, the uncertainty of the absorption cell temperature was maintained below 2 mK for the data analyzed in the present paper.

## III. LINESHAPE MODELS

The selected absorption line is the $\nu_2$ saQ(6,3) (J = 6 and K = 3 are respectively the quantum numbers associated with the total orbital angular momentum and its projection on the molecular symmetry axis) rovibrational line of the ammonia molecule $^{14}NH_3$ at the frequency $\nu$ = 28 953 693.9(1) MHz. This well-isolated transition has been chosen to avoid any deformation of the profile due to line mixing with neighbouring lines. Owing to the non-zero spin values of the nitrogen and hydrogen nuclei, an unresolved hyperfine structure is present in the Doppler profile of this rovibrational line. This hyperfine structure has been recorded by saturation spectroscopy to determine the induced broadening of the absorption linewidth recorded in linear absorption [24]. For a Doppler width of about 50 MHz, the linewidth broadening remains completely negligible (~0.25 kHz) for the present analysis of the linear absorption lineshape.

In this section, we give a brief description of semiclassical absorption lineshape models [28]. In the case of an optically thick medium, under low saturation, the Beer-Lambert law leads to:

$$P(L,\omega) = P_0 \exp\left[-\frac{4\pi^2 \alpha N d^2 \omega L e^{(-E_0/k_B T)}}{Z_{int}} I(\omega)\right] \quad (1)$$

where $P_0$ and $P(L,\omega)$ are respectively incident and transmitted light power, $\omega$ is the laser angular frequency, $E_0$ the energy of the lower rovibrational level in interaction with the laser electromagnetic field, $\alpha = e^2/(4\pi\varepsilon_0 \hbar c)$ the fine structure constant ($e$ is the electron charge, $\varepsilon_0$ the vacuum permittivity, $\hbar$ the reduced Planck constant and $c$ the velocity of light in vacuum), $N$ the density of molecules, $T$ the gas temperature, $L$ the absorption length, $d = \mu/e$ ($\mu$ is the transition moment) and $Z_{int}$ the internal partition function. The normalized absorption profile, $I(\omega)$, is related to molecular velocities and optical dipole relaxations.

At the very low pressure regime that we experimentally explore, the lineshape can simply be described by a Voigt profile (VP) which is the convolution of a Gaussian $I_D(\omega)$ (related to the inhomogeneous Doppler broadening) and a Lorentzian $I_L(\omega)$ (related to the homogeneous broadening due to molecular relaxation by collisions). Following notations of [29, 30]:

$$I_V(\omega) = \frac{1}{\sqrt{\pi}\,\Delta\omega_D}\,\text{Re}[w(x,y)] \quad (2)$$

with $x = \frac{\omega - \omega_0 - \Delta}{\Delta\omega_D}; y = \frac{\Gamma}{\Delta\omega_D}$ where $\omega_0$ is the angular frequency of the molecular line center, $\Delta$ the collisional shift, $\Gamma$ the width given by all homogeneous contributions to the broadening and $\Delta\omega_D = \omega_0 \sqrt{\frac{2k_B T}{m_{NH_3} c^2}}$ the profile Doppler width ($m_{NH_3}$ is the molecular mass).

$w(x,y) = \frac{i}{\pi}\int_{-\infty}^{+\infty} \frac{e^{-\tau^2}}{x - \tau + iy} d\tau$ is the complex probability function.

As will be shown later, this first approach is not sufficient to describe the details of the lineshape. With respect to the Voigt profile, the lineshape can be narrowed by either speed-dependent or Lamb-Dicke-Mössbauer (LDM) effects. We introduce three well-known models which take into account different physical cases leading to such narrowing. These models will be confronted to data in the next parts of the paper.

The dependence of the relaxation mechanism on molecular speed can be modelled by the speed-dependent Voigt profile (SDVP) where collisional shift and broadening, $\Delta(v)$ and $\Gamma(v)$, depend on the modulus of the molecular speed. Following the formalism developed by Berman, Ward and Pickett, the SDVP expression is given by [31-33]:

$$I_{SDV}(\omega) = \frac{1}{\pi} \text{Re} \left\{ \int \frac{f_M(\vec{v})}{\left[ \Gamma(v) - i(\omega - \omega_0 - \Delta(v) - \vec{k}.\vec{v}) \right]} d^3\vec{v} \right\} \quad (3)$$

where $f_M(\vec{v})$ is the Maxwell-Boltzmann distribution of molecular velocities $\vec{v}$ and $\vec{k}$ is the wave vector. The collisional parameters are:

$$\begin{cases} \Gamma(v) = \dfrac{\Gamma}{2^{m/2}} {}_1F_1\left( -\dfrac{m}{2}, \dfrac{3}{2}, -\left(\dfrac{v}{\bar{v}}\right)^2 \right) \\ \Delta(v) = \dfrac{\Delta}{2^{n/2}} {}_1F_1\left( -\dfrac{n}{2}, \dfrac{3}{2}, -\left(\dfrac{v}{\bar{v}}\right)^2 \right) \end{cases} \quad (4)$$

Where ${}_1F_1(a,b,z)$ is the Kummer confluent hypergeometric function, $\bar{v} = \sqrt{\dfrac{2k_B T}{m_{NH_3}}}$ is the most probable speed of molecules and $m$ and $n$ are parameters related to the intermolecular potential (in the particular case of $m = n = 0$, the homogeneous width and the collisional shift become independent of molecular velocities and formula (3) and (4) lead to a Voigt profile with $\Delta(v) = \Delta$ and $\Gamma(v) = \Gamma$).

When the mean-free path of molecules is much smaller than the laser wavelength, the LDM effect results in a reduction of the Doppler width [34]. Assuming soft collisions between molecules leads to the Galatry profile (GP) [35] which can be expressed using the ${}_1F_1$ function:

$$I_G(x,y,z_G) = \frac{1}{\pi \Delta\omega_D} \text{Re} \frac{1}{\left(-ix + y + \dfrac{1}{2z_G}\right)} {}_1F_1\left[ 1, 1 + \dfrac{-ix + y + \dfrac{1}{2z_G}}{z_G} ; \dfrac{1}{2z_G^2} \right] \quad (5)$$

with $z_G = \dfrac{\beta_G}{\Delta\omega_D}$ where $\beta_G$ represents the effective frequency of velocity changing collisions.

This profile evolves from a VP at low pressure to a Lorentzian in the high-pressure limit.

The LDM effect can also be modelled via the Rautian profile (RP) which assumes hard collisions between molecules (the velocity memory is lost after each collision) [36, 37]. The corresponding lineshape is given by:

$$I_R(x,y,z_R) = \frac{1}{\sqrt{\pi} \Delta\omega_D} \text{Re}\left\{ \frac{w(x, y+z_R)}{1 - \sqrt{\pi} z_R w(x, y+z_R)} \right\} \quad (6)$$

with $z_R = \dfrac{\beta_R}{\Delta\omega_D}$ and where $\beta_R$ is the effective frequency of velocity changing collisions.

In the frame of the GP (respectively RP), $\beta_{G(R)}$ is proportional to the gas pressure.

## IV. EXPERIMENTAL LINESHAPE ANALYSIS

Spectra have been recorded at different ammonia pressures, around 1.7 Pa in the MPC and from 9 to 20 Pa in the SPC. Linear absorption at resonance ranges between 83 % up to 98 % (see FIG. 2). Spectra, about 100 MHz wide, are recorded over 500 MHz by steps of 500 kHz with a 30 ms integration time per point. The time needed to record a single spectrum is about 70 s. For a proper adjustment of free parameters of the various lineshape models, the highest pressures are chosen so that total absorption is avoided and signal-to-noise ratio considerations set the low pressure limit. In these conditions signal-to-noise ratios as high as $10^3$ are reached for a single spectrum.

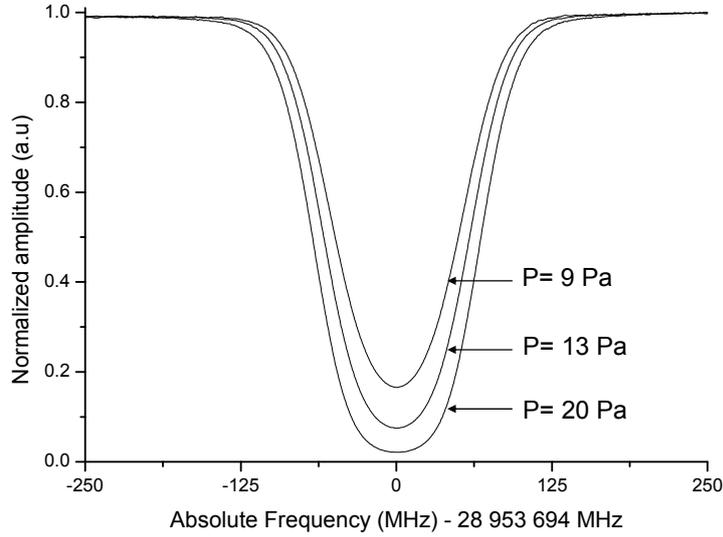

FIG. 2. Spectra recorded in SPC (single pass configuration) for pressures ranging from 9 to 20 Pa.

Experimental spectra have been analyzed by adding to the Beer-Lambert law, a baseline of slope $a_1$ as follows: $P(L,\omega) = \left[a_0 + a_1(\omega - \omega_0)\right] e^{\left[-A\,I(\omega)\right]}$ (7)
$A$ is the integrated absorbance (see equation (1)). Spectra of the self-broadened line of $NH_3$ have been fitted with a VP, SDVP, GP (Taylor expansion to second order in $z_G$ to reduce computing time) and RP using a least-squares-fit procedure. For these profiles common free parameters are $a_0$, $a_1$, $A$, $\Delta$, and $\Gamma$. $\omega_0$ has been deduced from high resolution saturated absorption spectra recorded in transmission of a Fabry-Perot cavity [24] and the Doppler width, $\Delta\omega_D$, is fixed to the value determined by the temperature of the gas. For the SDVP, two additional free parameters, $m$ and $n$, have to by fitted. For the GP (respectively RP), one additionnal parameter $\beta_{G(R)}$ has to be fitted. The amplitude noise is taken into account by weighting each individual point of each spectrum by considering the surrounding noise in the spectrum – this is directly related to the intensity noise on the photodetector which decreases strongly when absorption changes from 0 to 100%.

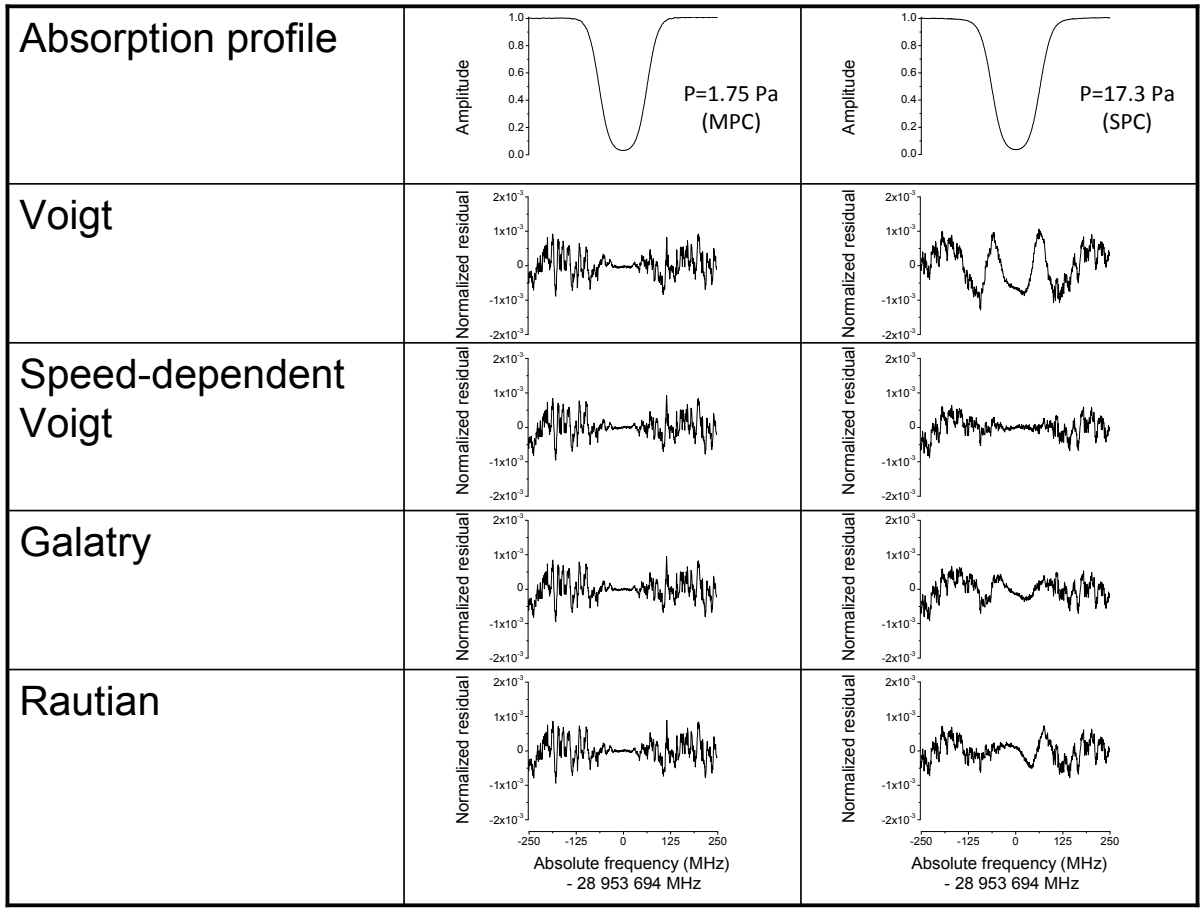

FIG. 3. Averaging of 24 spectra recorded at two pressures, 1.75 and 17.3 Pa and normalized residuals for non-linear least-squares fits, with a VP (Voigt Profile), SDVP (Speed-Dependent Voigt Profile), GP (Galatry profile) and RP (Rautian Profile) (rms noise amplitude is about 0.02 % of the absorption signal). The frequency scale is offset by 28 953 694 MHz.

The averaging of 24 spectra recorded at two $NH_3$ pressures, 1.75 and 17.3 Pa, is displayed on FIG. 3, together with residuals obtained after fitting data with a VP, SDVP, GP and RP. At 1.75 Pa all profiles considered reproduce equally well the recorded lineshape within the noise level. At the highest pressure of 17.3 Pa however, residuals for the VP, GP and RP clearly indicate that these profiles do not match the experimental data. The corresponding residuals are respectively 10.3, 4.3 and 6.7 times larger than the rms noise level. In FIG. 4 the frequency of velocity-changing collisions determined from fitting individual spectra with a GP and RP are plotted as a function of pressure. Polynomial interpolation of this data reveals in both cases a non-linear dependence of $\beta_{G(R)}$ with ammonia density (clearly confirmed by an F-test with 5% of confidence level).

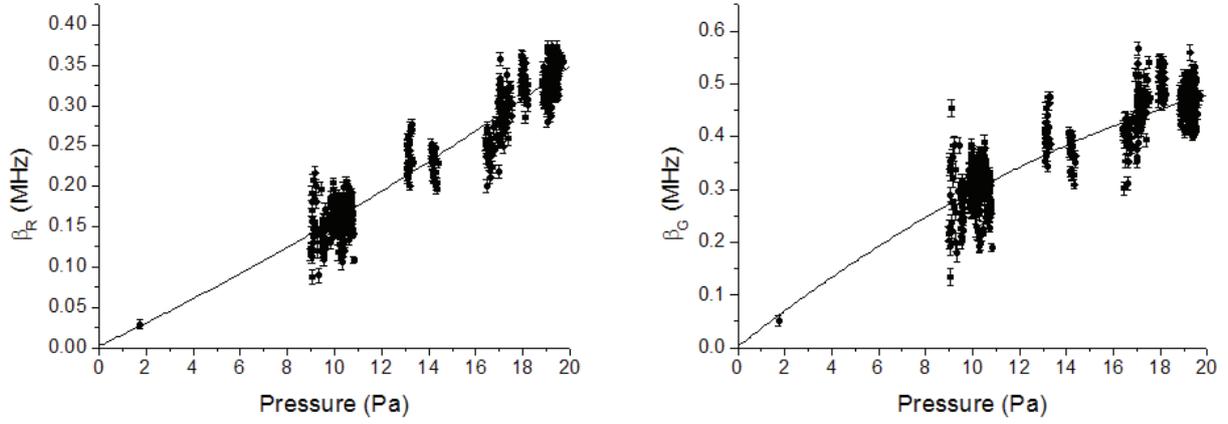

FIG. 4. Frequency of velocity-changing collision parameters as a function of ammonia pressure, deduced from fitting individual spectra with a GP (Galatry Profile) or RP (Rautian Profile). Associated 2$^{nd}$ order polynomial fits are shown. An F-test (with 5% of confidence level) indicates a non zero 2$^{nd}$ order polynomial coefficient and higher order coefficients consistent with zero.

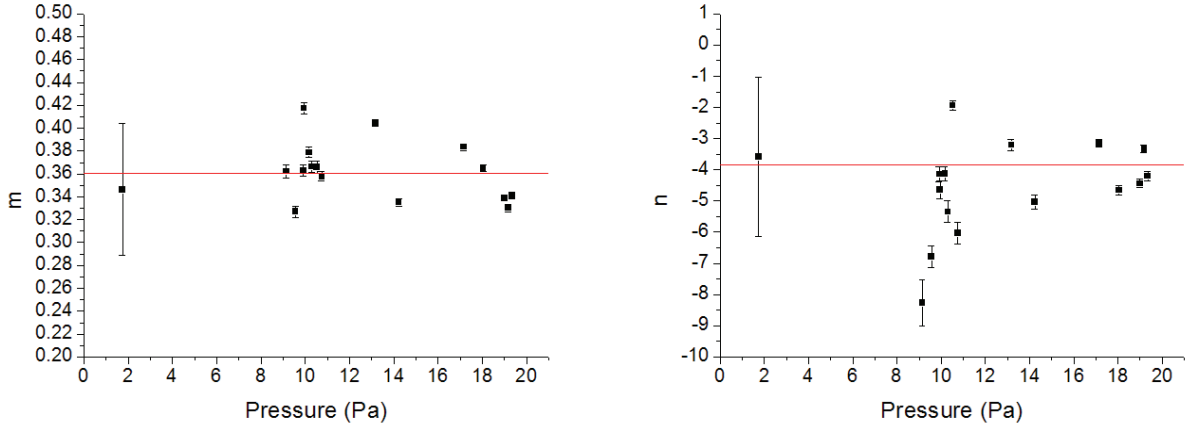

FIG. 5. *m* and *n* parameters as a function of pressure retrieved from fitting data (averaging of 24 individual spectra) with the SDVP (Speed-Dependent Voigt Profile). The discrepancy between the standard deviation estimated from the dispersion of *m* and *n* and the error bar of each point (deduced from the fit) is attributed to long term instabilities of the spectrometer.

Results reported in FIG. 3 and FIG. 4 led to reject the VP, GP as well as RP for the modelisation of self-broadening of the $NH_3$ line over the investigated pressure range. Furthermore the GP and RP residuals clearly show an asymmetry which is a good indication of a speed dependence of the pressure shift. By contrast, analysis of spectra with the SDVP clearly demonstrates a good agreement between this model and experimental lineshapes at 17.3 Pa. In FIG. 5 the determined *m* and *n* parameters are plotted as a function of pressure. As expected these parameters are independent of the molecular density. Based on our analysis, we can state that the deviation from a VP of the $\nu_2$ saQ(6,3) line of $NH_3$, in the covered pressure range, is most probably caused by speed-dependent effects and not LDM effects (that induce a reduction of the Doppler width).

In TABLE I fitted lineshape parameters obtained using the VP, SDVP, GP and RP are summarized. The retreived homogeneous broadening as well as collisional shift parameters are consistent for models that take into account narrowing effects (the SDVP, GP and RP). On the other hand a large discrepancy (up to 10%, for pressures ranging from 1.7 to 20 Pa) is observed on the homogeneous broadening parameter obtained by fitting a Voigt profile to experimental data. Note that such discrepancies for the self-broadening parameter have also been reported at higher pressure in the R branch of the $\nu_2$ band by Guinet *et al.* [19]. In the assumption of an intermolecular potential given by $V(r) \propto \frac{1}{r^q}$ (with *r* the intermolecular distance), *m* parameter is related to the power law by $m = \frac{q-3}{q-1}$. The retrieved *m* value equal to 0.360(7) would correspond to *q* close to 4 (a dipole-quadrupole interaction potential). Following the formalism developed by Berman and Ward [31, 32] *n=m* and *n* = −3.8(3) would correspond to *q* close to 1.4. Another approach developed by Pickett [33] leads to $n = \frac{-3}{q-1}$ for which *n* parameter would correspond to *q* close to 1.8. Such inconsistencies between *m* and *n* have already been observed by several authors and could be an indication of a more complicated intermolecular potential [28, 38, 39]. A physical interpretation of these coefficients remains thus questionable but is beyond the scope of the present work.

| PARAMETERS | LINESHAPE PROFILE | | | |
|---|---|---|---|---|
| | Voigt | Speed-dependent Voigt | Galatry | Rautian |
| $d\Gamma/dp$ (kHz/Pa) | from 106 to 130 (from 0 to 20 Pa) | 120(3) | 120.9(3) | 119.9(3) |
| $d\Delta/dp$ (kHz/Pa) | 0.99(4) | 1.2(1) | 1.01(4) | 1.04(4) |
| *m* | | 0.360(7) | | |
| *n* | | -3.8(3) | | |
| $d\beta/dP$ (kHz/Pa) | | | from 32 to 14 (from 0 to 20 Pa) | from 14 to 20 (from 0 to 20 Pa) |

TABLE I. Lineshape spectroscopic parameters (and their standard uncertainties) derived from various line profiles for the self-broadened $\nu_2$ saQ(6,3) line of $NH_3$ around 273.15 K.

### V. APPLICATION TO THE DETERMINATION OF THE BOLTZMANN CONSTANT

Influence of the line-shape model on the spectroscopic determination of the Boltzmann constant has also already been raised by us and other groups for $O_2$ and $H_2O$ [24, 27, 38-42]. The interest in this constant is related to the forthcoming redefinition of the International System of Units (expected to happen at the next meeting of the Conférence Générale des Poids et Mesures). A new definition of the kelvin would fix $k_B$ to a value determined by The Committee on Data for Science and Technology (CODATA) [43-45]. In this section, we exploit the accurate knowledge of the absorption lineshape gained in section IV. to reduce the main source of uncertainty in the Boltzmann constant measurement by laser spectroscopy.

The spectrometer presented in section II. is part of an experimental setup dedicated to the measurement of the Boltzmann constant. We measure $k_B$ by the determination of the Doppler width (proportional to $\sqrt{k_B T}$) of the $\nu_2$ saQ(6,3) ammonia rovibrational line, in thermal and pressure equilibrium [23, 46-50] (other groups have recently started experiments on $CO_2$, $H_2O$, acetylene and rubidium [51-56]). We have lately reported experimental developments aiming at reducing the uncertainty on $k_B$ [24, 26, 27]. The current uncertainty budget shows a combined standard uncertainty of 144 ppm, the main contribution coming at 97% from the lack of knowledge in the collision modelisation and the appropriate resulting lineshape. Since we have clearly identified in section IV. the SDVP as the best lineshape profile, the uncertainty coming from collisional effects is now only limited by our knowledge on the associated parameters $m$ and $n$. From data shown on FIG. 5, these parameters can be estimated with an uncertainty of 2.5% for $m$ and 8% for $n$. The aim of our project being an independent determination of $k_B$, these uncertainties of $m$ and $n$ take into account an overestimated uncertainty of 100 ppm for $k_B$ (that given by the CODATA is only 1.7 ppm) corresponding to an uncertainty of 50 ppm for the fixed value of $\Delta\omega_D$.

In order to reduce the impact of the uncertainty associated with these two collisional parameters on the lineshape, the measurement of the Boltzmann constant is performed at low pressure, between 1 and 2 Pa in the MPC. At these pressures the signal to noise ratio remains high enough for a numerical adjustment of 6 free parameters of the SDVP: $a_0$, $a_1$, $\Delta$, $A$, $\Gamma$ and $\Delta\omega_D$ the Doppler width from which is deduced $k_B$ ($m$, $n$ and $\omega_0$ are then fixed). New series of spectra will be recorded over a frequency scale of 500 MHz under well controlled experimental conditions (laser frequency, laser intensity, gas temperature, gas purity,...). A high purity of the gas sample will be ensured by a systematic control of the residual pressure and outgasing of the absorption cell. Simulated spectra corresponding to these experimental conditions have been fitted taking into account uncertainties on $m$ and $n$ given in TABLE I in order to estimate the impact of these uncertainties on the Doppler width accuracy. We followed the line-absorbance based analysis method recently proposed by A. Castrillo *et al.*, expected to be robust with respect to the lineshape modelisation [57]. In this approach, which uses the relationship between the line-center absorbance and the integrated absorbance, the Doppler width is obtained from a global analysis of spectra recorded at different pressures. The experimental uncertainties obtained for $m$ and $n$ lead to an uncertainty of 1.8 ppm on the Doppler width determination. The present work clearly indicates that a first determination of the Boltzmann constant by laser spectroscopy with an uncertainty of a few ppm is now reachable.

## VI. CONCLUSION

We have reported on an accurate analysis of self-broadened $\nu_2$ saQ(6,3) lineshape of $NH_3$ around 273.15 K which show for the first time a clear evidence of speed-dependent broadening and shifting. Accurate lineshape parameters have been obtained considering a speed-dependent Voigt profile. From the accurate knowledge of the absorption lineshape we anticipate a reduction the main source of uncertainty in the Boltzmann constant measurement by laser spectroscopy. The uncertainty coming from collisional effects is now only limited by our knowledge on the associated parameters $m$ and $n$ and we conclude that a first determination of $k_B$ with a competitive uncertainty of a few ppm by measurement of the Doppler width in ammonia is now reachable. It will then be worthily compared to the value obtained by the acoustic method and thus hopefully contribute significantly to the CODATA value.

## AKNOWLEDGMENTS

This work is funded by CNRS, the Laboratoire National de Métrologie et d'Essais and by European Community (EraNet/IMERA). Authors would like to thank Y. Hermier, F. Sparasci and L. Pitre from Laboratoire Commun de Métrologie LNE-CNAM for platinum resistance thermometers calibrations, S. Briaudeau for helping in the design and setting up of the thermostat and temperature control devices. We are indebted to Pr. Ch. J. Bordé for having originally proposed this novel approach for measuring the Boltzmann constant by laser spectroscopy and for many fruitful discussions over the course of this project.